\documentclass[aps,pra,twocolumn,superscriptaddress]{revtex4-1}

\usepackage{graphicx}

\begin{document}


\title{Understanding photodetector nonlinearity in dual-comb interferometry}


\author{Philippe Guay}
\email[]{philippe.guay.4@ulaval.ca}
\affiliation{Centre d'optique, photonique et laser, Universit\'{e} Laval, Qu\'{e}bec, Qu\'{e}bec G1V 0A6, Canada}

\author{Alex Tourigny-Plante}
\affiliation{Centre d'optique, photonique et laser, Universit\'{e} Laval, Qu\'{e}bec, Qu\'{e}bec G1V 0A6, Canada}

\author{Vincent Michaud-Belleau}
\affiliation{Centre d'optique, photonique et laser, Universit\'{e} Laval, Qu\'{e}bec, Qu\'{e}bec G1V 0A6, Canada}
\affiliation{Now with LR Tech inc., Lévis, QC G6W 1H6, Canada}

\author{Nicolas Bourbeau Hébert}
\affiliation{Centre d'optique, photonique et laser, Universit\'{e} Laval, Qu\'{e}bec, Qu\'{e}bec G1V 0A6, Canada}
\affiliation{Now with Institute for Photonics and Advanced Sensing (IPAS) and School of Physical Sciences, University of Adelaide, Adelaide, SA 5005, Australia}

\author{Ariane Gouin}
\affiliation{Centre d'optique, photonique et laser, Universit\'{e} Laval, Qu\'{e}bec, Qu\'{e}bec G1V 0A6, Canada}

\author{J\'{e}r\^{o}me Genest}
\affiliation{Centre d'optique, photonique et laser, Universit\'{e} Laval, Qu\'{e}bec, Qu\'{e}bec G1V 0A6, Canada}



\date{\today}

\begin{abstract}
The impact of photodetector nonlinearity on dual-comb spectrometers is described and compared to that of Michelson-based Fourier transform spectrometers (FTS). The optical sampling occurring in the dual-comb approach, being the key difference with FTS, causes optical aliasing of the nonlinear spectral artifacts. Measured linear and nonlinear interferograms are presented to validate the model. Absorption lines of H$^{13}$CN are provided to understand the impact of nonlinearity on spectroscopic measurements. 
\end{abstract}


\maketitle

\section{Introduction\label{intro}}
The impact of the nonlinear response of a photodetector has been extensively studied with conventional Fourier transform spectrometers (FTS) \cite{LAC00,CHA84,GUE86,CAR90,ABR94,JES98}. However, its impact on dual-comb spectrometers (DCS) has yet to be fully investigated. FTS and DCS are similar spectroscopic approaches since both methods yield an interferogram (IGM) related to field correlation functions. In the FTS case,  the IGM results from the beating of a source signal with a delayed version of itself (autocorrelation) while in DCS, the IGM results from the beat of two pulsed lasers with detuned repetitions rate (cross-correlation). Since the dual-comb IGM is the result of the interference between two pulse trains, one pulse train samples the other in a way that is similar to equivalent-time sampling used by digital oscilloscopes. This operation known as optical sampling is a key difference between the two approaches. As a result of optical sampling, the dual-comb spectrum becomes periodic and photodetector (PD) nonlinearity (NL) manifests itself differently. An understanding of nonlinearity and the generation of spectral artifacts is however necessary before discussing the impact of optical sampling. 

\begin{figure}[htbp]
\includegraphics[width=0.5\textwidth]{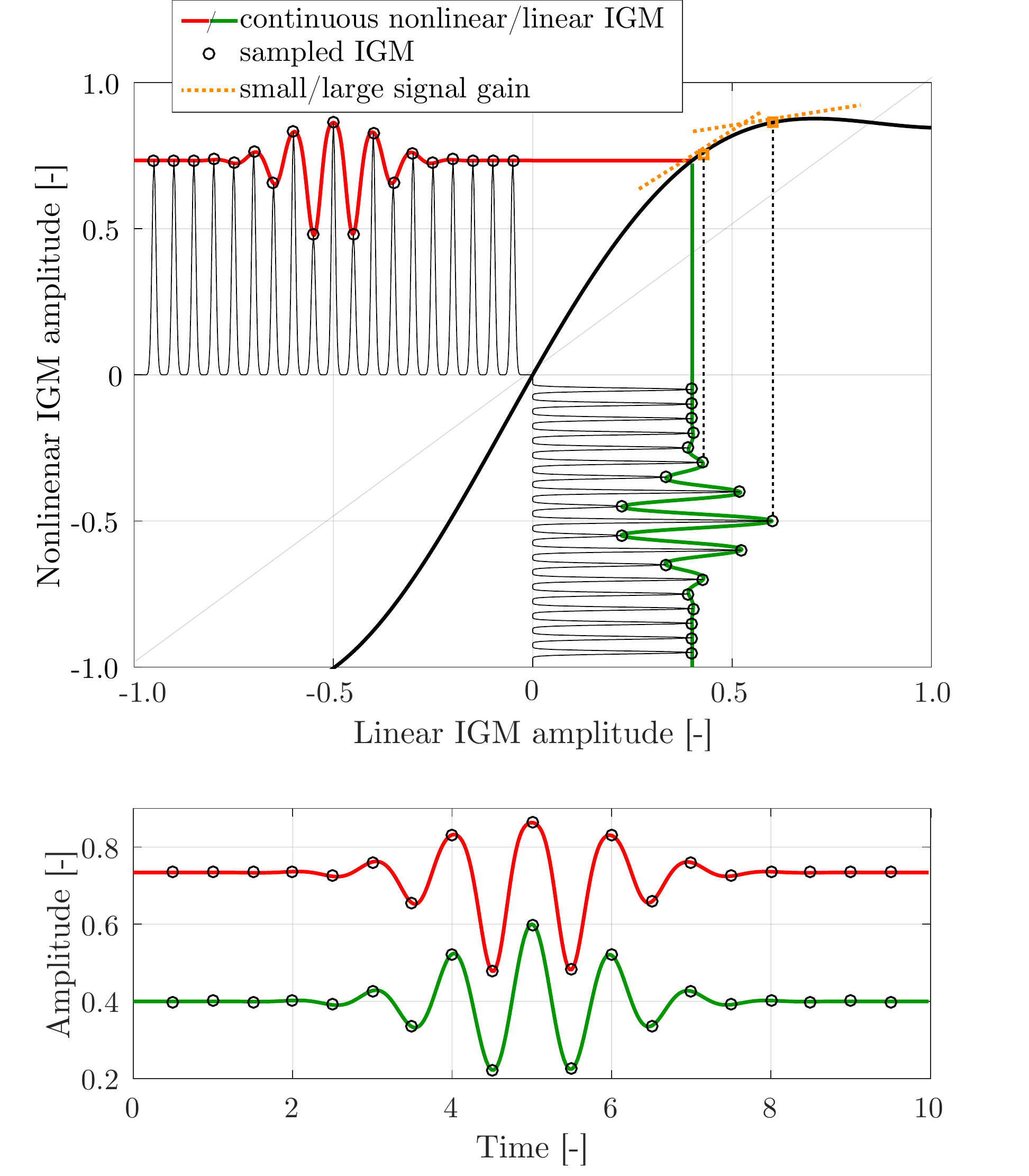}
\caption{\label{fig:NLconv} Nonlinear conversion of an interferogram for the continuous and sampled cases.}
\end{figure}

One key hypothesis in classical FTS NL studies is that NL is assumed to be static. That is to say there is, as shown in the top panel of Fig. \ref{fig:NLconv}, a one to one relation between the linear and nonlinear interferograms. It is supposed that there is no dynamic or memory effects occurring in the process. A continuous linear IGM (green in Fig.  \ref{fig:NLconv}) is transformed at any given time into the measured nonlinear IGM (red in Fig. \ref{fig:NLconv}).

Optical sampling in DCS occurs in the detector and is thus intrinsically linked to nonlinearity. Depending upon where nonlinearity occurs, for instance in the photodiode or in the amplifying chain, it can be conceptualized as occurring during or after optical sampling. Fig. \ref{fig:NLconv} shows that inverting the order of theses operations produces similar results providing there is no in-between bandwidth limiting filters. The pulses on the continuous IGMs on Fig. \ref{fig:NLconv} illustrate the detector's impulse response sampling the IGMs. Provided that the detector's impulse response stays the same, sampling the linear IGM and applying the static NL transformation shown on on Fig. \ref{fig:NLconv} yields the same result as carrying the NL transformation on the continuous signal and sampling afterwards. To some extent, changes in the sampling function properties can be expressed with a constant sampling function and a modified continuous NL relation. 

\begin{figure}[htbp]
\includegraphics[width=0.5\textwidth]{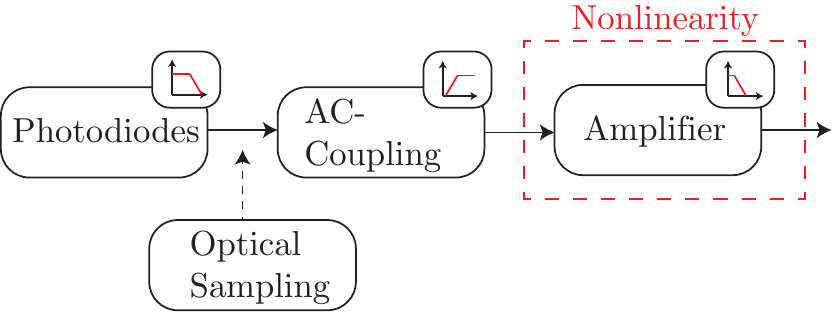}
\caption{\label{fig:order} Assumed order of operations in the photodetection chain.}
\end{figure}

The order in which optical sampling and nonlinearity occur in the detector may not be relevant here, but it is worth describing the overall order of operations occurring in the photodetection process. In amplified balanced detectors commonly used in DCS experiments, it was demonstrated \cite{GUA21a} that NL occurs mainly in the final amplification step, where the second operational amplifier is close to saturation. So, the operations assumed here, including the current generation in the diodes, the AC-coupling step often used in detectors for DCS experiment and the amplification, are shown in Fig. \ref{fig:order}. The exact location where nonlinearity occurs in the detection chain and the presence of any filter in the chain will influence how nonlinearity arises. If nonlinearity happens before a filter whose impulse response spreads across pulses, a dynamic description of NL becomes required. Here, the photodiodes and AC-coupling steps are occurring before the nonlinearity generation in the amplifier. Moreover, as the bandwidths of the diode, the AC-coupler, and the amplifier are chosen to be much larger than the repetition rate of the laser, the detector's impulse responses are well separated \cite{GUA21a,DID09} and do not overlap one another. As a result, dynamic NL effects are minimized.

Pursuing with the static NL hypothesis, the transformation between the linear and measured nonlinear IGM can be written as a polynomial expansion :

\begin{equation}\label{poly}
\text{IGM}_{\text{NL}} =  a_0 + a_1[\text{IGM}_{\text{L}}]  + a_2[\text{IGM}_{\text{L}}]^2  + ...
\end{equation}

where $\text{IGM}_{\text{NL}}$ and $\text{IGM}_{\text{L}}$ are respectively the nonlinear and linear interferograms and where $a_0$, $a_1$ and $a_2$ are respectively the constant coefficient, the linear coefficient and the NL coefficient of second order. Since the interferogram ($I_m$) is often measured with an AC-coupled detector, it is convenient to redefine the NL transformation in terms of a zero-mean linear interferogram (IGM$_{\text{0L}}$). This yields a new series expansion \cite{ABR94} that simply amounts to a redefinition of the x-axis in Fig.\ref{fig:NLconv} such that the point around which the series is expanded is labelled as zero:

\begin{equation}
I_m = A_0 + A_1 [\text{IGM}_{\text{0L}}] + A_2[\text{IGM}_{\text{0L}}]^2+...
\end{equation}

Redefining the y-axis in Fig. 1 such that the measured IGM is also zero-mean allows us to pose $A_0=0$. Another way to write equation 2 that provides insight with $A_0=0$ and factorising the linear IGM is:

\begin{equation}
I_m = \text{IGM}_{\text{0L}} ( A_1 + A_2 \text{IGM}_{\text{0L}} + …)
\end{equation}

With this notation, the term in parenthesis acts as a “gain” for the linear IGM. The linear coefficient $A_1$ acts as a constant gain that usually reduces the level of the linear signal, a consequence of the generation of higher harmonics. This leads to a signal-to-noise degradation for the frequencies of interest, but this shall provide no systematic error upon calibration for instance in a transmittance measurement, provided the signal level is similar for the reference measurement. Subsequent terms provide a gain that varies with signal strength. This is readily apparent on the second order term, where the “gain” on $\text{IGM}_{\text{0L}}$ varies with $\text{IGM}_{\text{0L}}$. This means that the baseline low resolution spectrum linked to the shape of the IGM peak through the Fourier transform does not experience the same gain as small signals in the wings of the IGM, such as the free induction decay caused my molecular absorption features. This can be seen on Fig. \ref{fig:NLconv} where the small and large signal gains are identified to be different

\begin{figure}[htbp]
\includegraphics[width=0.5\textwidth]{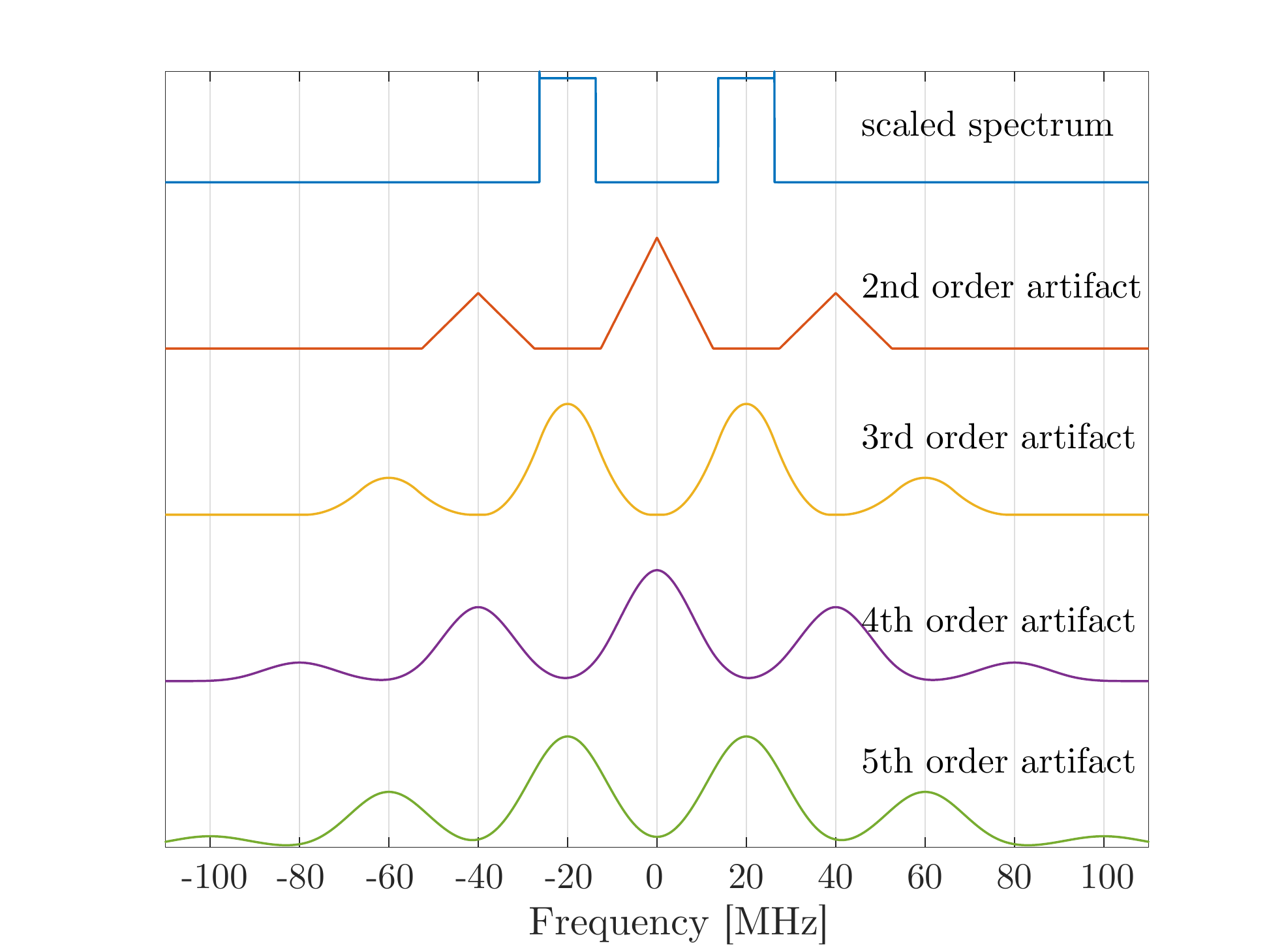}
\caption{\label{fig:artifacts} Spectral artifacts up to the fifth order generated from a nonlinear measurement of the uniform spectral distribution shown in blue.}
\end{figure}

In the spectral domain, the n-th nonlinearity order is construed as the (n-1)-times convolution of the spectrum with itself. For instance, the second order nonlinear term is the auto-convolution of the spectrum. Fig. \ref{fig:artifacts} illustrates the spectral artifacts for the first 5 orders for a uniform spectral distribution generated around 1$f$~=~20~MHz.  The second order term has a spectral contribution at twice the frequency of the signal (2$f$), but also around DC. The third order term has a spectral contribution at three times the frequency (3$f$) of the linear signal, but also generates content overlapping at $1f$. This process can be generalized to any NL order.

It is worth emphasizing that applying a band-pass filter at $1f$ is not sufficient to get a linear signal. As explained above, nonlinear artifacts create a gain that varies with signal strength  so features having different widths translating to different spreads across the IGM will experience a different gain. Moreover, odd NL artifacts generate spectral content that additively combine in the band of interest, a definitive source of irregular systematic spectral errors. 

\begin{figure}[htbp]
\includegraphics[width=0.5\textwidth]{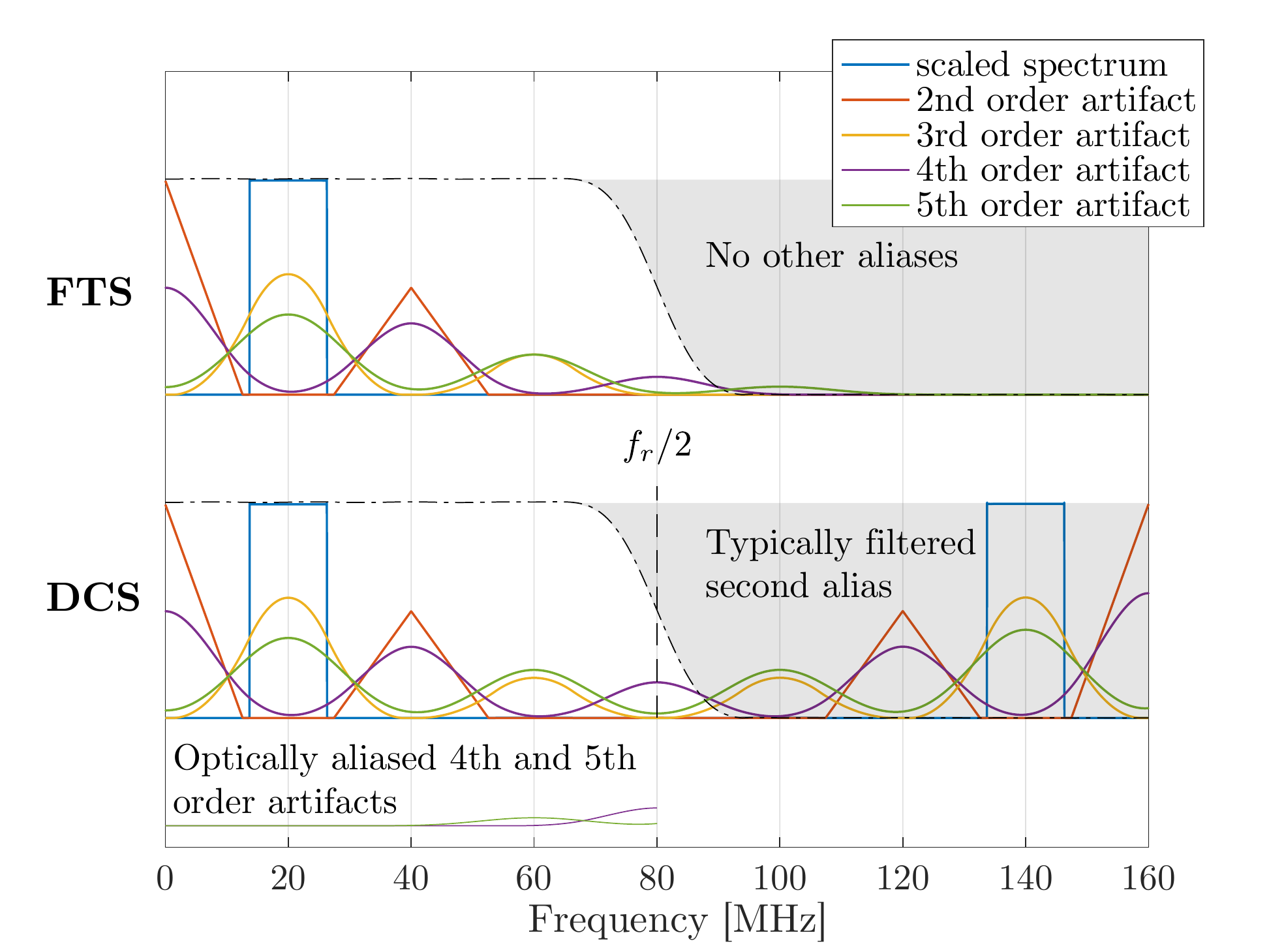}
\caption{\label{fig:FTSversusDCS} Spectrum showing nonlinear artifacts up to the fifth order for the case of a Fourier transform spectrometer (FTS) and a dual-comb spectrometer (DCS) where the main difference lies in the optical aliasing of the fourth and fifth order. }
\end{figure}

Now that the NL generation of spectral artifacts has been described, the effect of optical sampling can be better understood. Sampling the interferogram periodises the spectrum and creates a spectral alias at each repetition rate $(f_r)$ multiple, limited by the bandwidth of the sampling function. Whether the optical sampling occurs before or after the nonlinearity, the result is similar given the assumptions made here: the spectrum is periodised and NL artifacts are aliased. The resulting spectrum for the case of a FTS IGM is shown in the top of Fig. \ref{fig:FTSversusDCS} while the  IGM is shown in bottom of the same figure for the DCS case where the repetition rate of the two lasers is 160 MHz.  In that case, the fourth and fifth order spectral artifacts are aliased as their spectral location exceeds half of the lasers' rep rates. One can see this as folding the spectral information that is found above 80 MHz. This explains why the green curve no longer overlaps the yellow one at 60~MHz and why the spectral fourth order artifact at 80~MHz becomes twice as important.

Even if only the first alias of a DCS measurement is kept by filtering the signal above half the repetition rate of the laser, optical aliasing has occurred and the fourth and fifth order term have already been summed with the signal. One can see that this can become problematic when the $1f$ signal of interest is spread over a large fraction of the  $f_r/2$ band and that all the nonlinear artifacts are folded and overlapped. The presented example is simplified in a sense that the artifacts are mostly separated and that only the fourth and the fifth order are aliased.

\section{Experimental methods}

In order to experimentally observe the effect of static nonlinearity coupled with optical sampling, the bandwidth of the detector was chosen to allow separation of the detector's impulse responses. Moreover, the repetition rate difference between the two frequency combs was precisely tuned to wisely place the nonlinear artifacts in a configuration that minimizes optical aliasing and overlap. This condition is the one presented in Fig. \ref{fig:FTSversusDCS}. The objective here was to observe NL spectral artifacts to validate the model. By doing so, the experiment brings insights into the impact of photodetector nonlinearity on dual-comb experiments. 

\begin{figure}[htbp]
\includegraphics[width=0.5\textwidth]{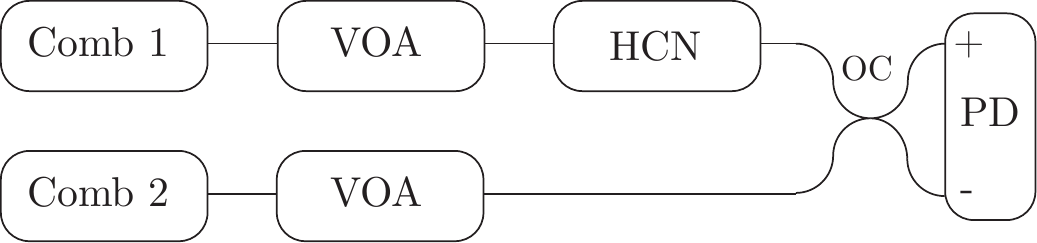}
\caption{\label{fig:expsetup} Simplified block diagram of the experimental setup. VOA: Variable optical attenuator. OC: 50/50 optical coupler. PD: Photodetector. HCN : H$^{13}$CN gas filled cell}
\end{figure}

The setup is shown on Fig. \ref{fig:expsetup} where two custom-made passively mode-locked lasers based on an erbium-doped fiber were used \cite{SIN15}. The lasers' central wavelength is 1550~nm and their repetition rate is 160~MHz. The lasers followed by an optical variable attenuator were combined by a 50/50 optical coupler and sent on a balanced photodetector (Thorlabs PDB480C). The lasers repetitions rates were adjusted so that the repetition rate difference is about 150~Hz. The carrier-offset frequency of the lasers ($f_{\text{CEO}}$) was adjusted so that the signal of interest is centered at 20~MHz. Interferograms were measured for low intensity pulses as well as for a signal clearly saturating the detector at the IGM centerburst.

A gas cell filled with hydrogen cyanide H$^{13}$CN was added in one arm before the output coupler. 


Due to the requirement of high sampling rate, but also because of memory limitation, only the central portions of the interferograms have been acquired \cite{TOU20}. Consequently, spectral resolution is limited, but since the need of this experiment requires only a relative comparison between linear and nonlinear measurements, this does not affect the interpretation of the results. 
To properly assess the impact of NL on the spectral region of interest here, the absorption lines of H$^{13}$CN for the two datasets shown in Fig. 6 are compared. For spectral transmittances, 80 high power (50~$\mu$W) and 1000 low power (10~$\mu$W) IGMs were digitized at their centerburst (1 ms at ZPD). Once digitized, the IGMs have been aligned, phase-corrected \cite{HEB17,GUA18} and averaged to limit the noise at a level allowing to clearly distinguish the impact of NL. The spectrum is normalized by fitting a eighth-order polynomial to the smooth spectral baseline instead of using an independent reference measurement \cite{GUA20}.

\section{Results}

\begin{figure}[htbp]
\includegraphics[width=0.5\textwidth]{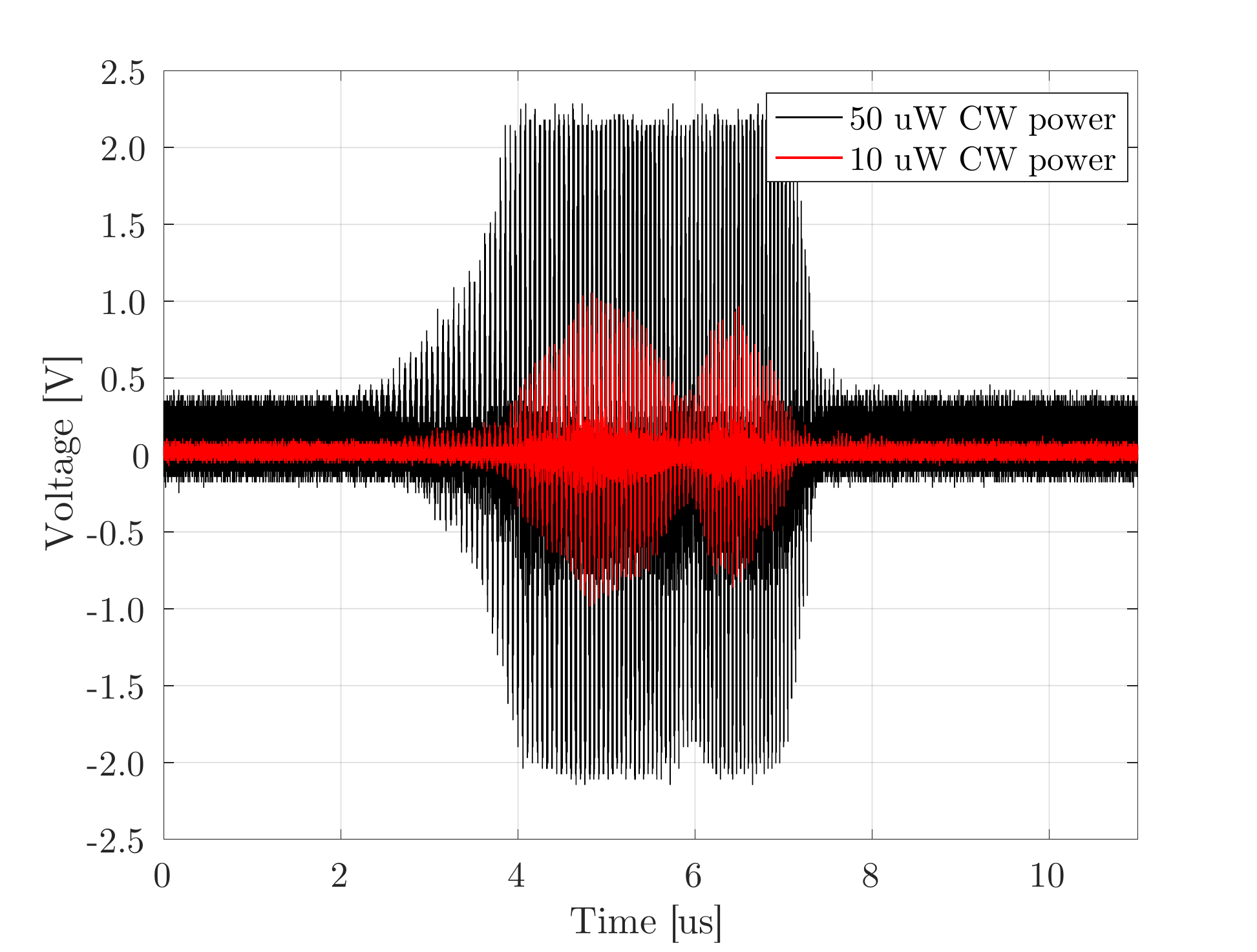}
\caption{\label{fig:IGMs} Linear low power (red) and nonlinear high power (black) interferograms. }
\end{figure}

Linear and nonlinear interferograms were measured having respectively 10 $\mu$W and 50 $\mu$W average powers at the photodetector. The waveforms are shown in Fig. \ref{fig:IGMs}. It can readily be seen that the nonlinear IGM saturates the amplified photodetector to its rail of $\pm$2 V. It is worth noting that 50 $\mu$W is sufficient to saturate the photodetector's amplifier even though its CW saturation level is nearly an order of magnitude higher at 400 $\mu$W. This is explained by the fact that short pulses have a much greater peak power for a given CW equivalent power. 

\begin{figure}[htbp]
\includegraphics[width=0.5\textwidth]{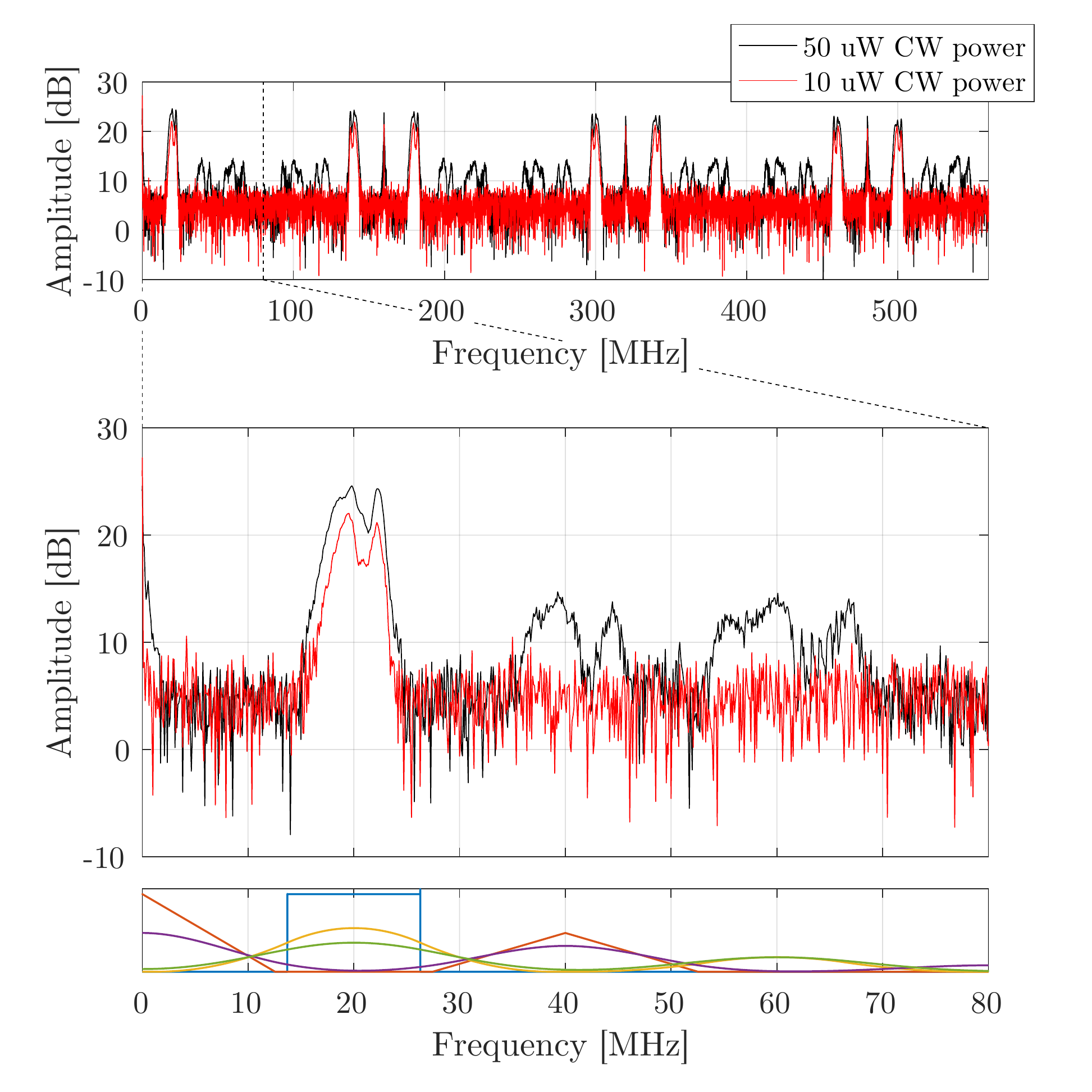}
\caption{\label{fig:spectrum} Spectrum of the low power (red) and high power (black) interferograms showing nonlinear artifacts and their expected position at the bottom of the plot. }
\end{figure}

The spectra of the low power and high power IGMs are shown in Fig. \ref{fig:spectrum} where many aliases are displayed in the top panel to highlight the periodisation of spectral artefacts. Only the central portion of the IGMs as shown on Fig. \ref{fig:IGMs} has been used to compute the spectrum. This allows a better visualization of the spectral artifacts since most of the nonlinearity occurs near zero path difference (ZPD) and since reducing the observation window reduces the noise floor level. The middle panel focuses on the first alias where the low intensity interferogram produces spectral content centered around 20 MHz and is otherwise limited by the oscilloscope additive noise. The high power measurement shows nonlinear artifacts that are generated at the expected locations. The second order term is visible at DC and at 40 MHz, while the third order nonlinearity is visible at 60 MHz and is known to also have a contribution at 20 MHz. The fourth and fifth order are not clearly not visible and  might not have a sufficient SNR to stand out of the noise.

\begin{figure}[htbp]
\includegraphics[width=0.5\textwidth]{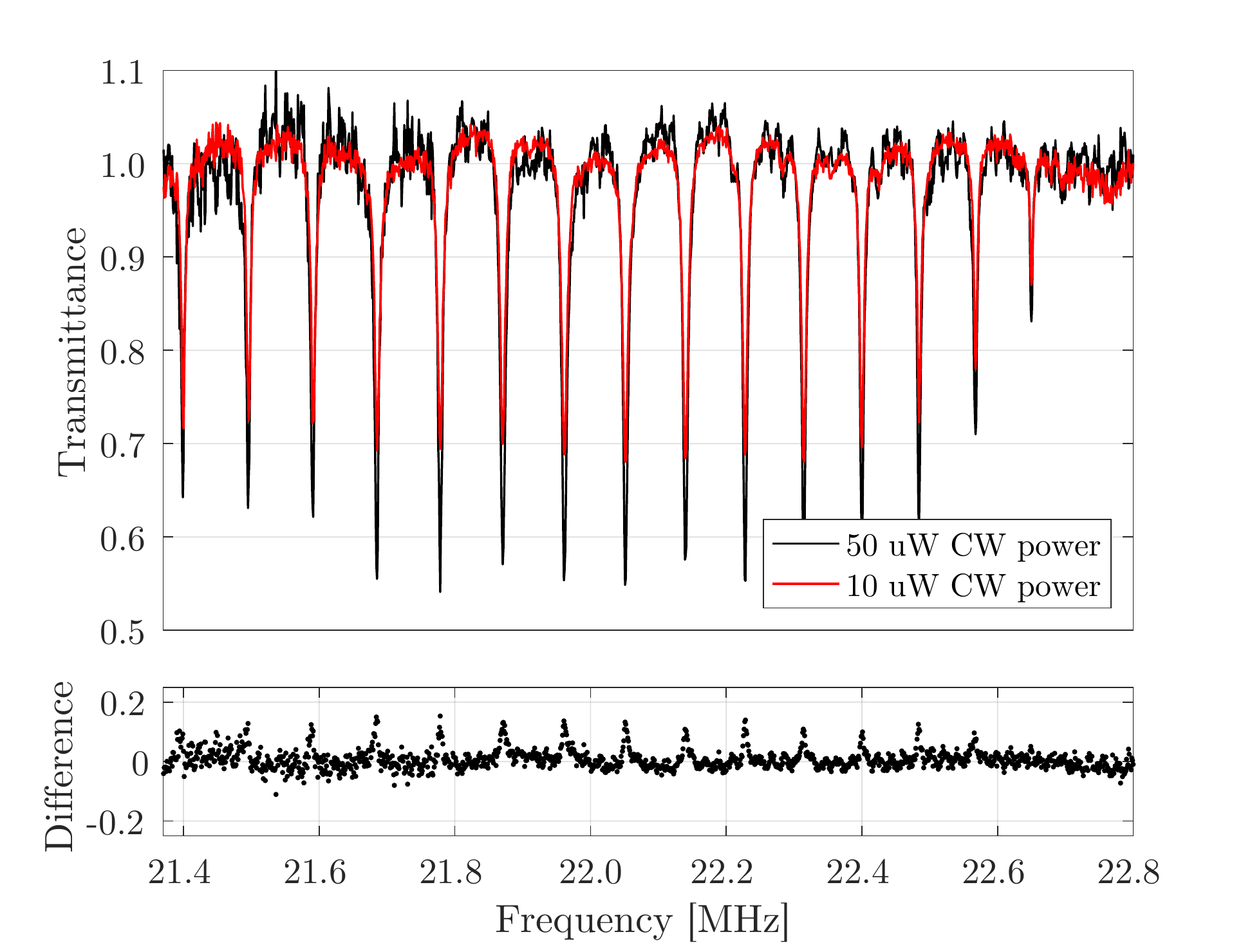}
\caption{\label{fig:spectro} Transmittance spectrum (top panel) of H$^{13}$CN for the P15 to P1 lines in the case of a high power nonlinear measurement and a linear measurement and the difference between the measurements (bottom panel). }
\end{figure}

In order to better evaluate the distortions produced on a spectroscopic measurement, absorption lines of H$^{13}$CN are measured. On Fig. \ref{fig:spectro}, it can be seen that the depth of the lines is greater for the high power measurement. This is a distortion of the signal that is explained by nonlinearity: as the IGM saturates, the IGM portion at ZPD, mostly the baseline of the spectrum, is mostly affected while the small signal such as the free induction decay in the wings of the IGM remains mostly intact. This translates to a baseline level underestimated for the same absorption lines strength and, thus absorption lines get deeper when normalized to provide transmittance data.

It is worth mentioning that the depth of the absorption lines may not match line-resolved spectroscopic analysis here \cite{LAR20} for the gas cell parameters as having measured only the IGM centerbursts with a single point phase correction induces an instrument lineshape (ILS) that reduces the depth of the lines \cite{POT13}. Nevertheless, as the ILS is the same for both measurements, the comparison remains valid. 


\section{Conclusion}

As a conclusion, a greater understanding of the impact of photodetector nonlinearity has been provided for dual-comb spectroscopy. A model of static NL  used in classical FTS has been adapted for DCS where differences mostly due to optical sampling were highlighted. Experimental data in a configuration of clear nonlinearity display was provided to support the model and further the understanding of nonlinearity on spectroscopic measurements. The nonlinearity distorts the interferogram and the spectroscopic lines were deepened, resulting in incorrect line intensities. With a proper NL model, these systematic errors can be taken into account properly in the data processing chain.

\begin{acknowledgments}
This work was supported by Natural Sciences and Engineering Research Council of Canada (NSERC),  Fonds de Recherche du Québec - Nature et Technologies (FRQNT) and the Office of Sponsored Research (OSR) at King Abdullah University of Science and Technology (KAUST) via the Competitive Research Grant (CRG) program with grant \# OSR-CRG2019-4046

The authors thank Ian Coddington at NIST for providing the dual-comb system.
\end{acknowledgments}

\bibliography{apstemplate.bib}

\begin{thebibliography}{15}%
\makeatletter
\providecommand \@ifxundefined [1]{%
 \@ifx{#1\undefined}
}%
\providecommand \@ifnum [1]{%
 \ifnum #1\expandafter \@firstoftwo
 \else \expandafter \@secondoftwo
 \fi
}%
\providecommand \@ifx [1]{%
 \ifx #1\expandafter \@firstoftwo
 \else \expandafter \@secondoftwo
 \fi
}%
\providecommand \natexlab [1]{#1}%
\providecommand \enquote  [1]{``#1''}%
\providecommand \bibnamefont  [1]{#1}%
\providecommand \bibfnamefont [1]{#1}%
\providecommand \citenamefont [1]{#1}%
\providecommand \href@noop [0]{\@secondoftwo}%
\providecommand \href [0]{\begingroup \@sanitize@url \@href}%
\providecommand \@href[1]{\@@startlink{#1}\@@href}%
\providecommand \@@href[1]{\endgroup#1\@@endlink}%
\providecommand \@sanitize@url [0]{\catcode `\\12\catcode `\$12\catcode
  `\&12\catcode `\#12\catcode `\^12\catcode `\_12\catcode `\%12\relax}%
\providecommand \@@startlink[1]{}%
\providecommand \@@endlink[0]{}%
\providecommand \url  [0]{\begingroup\@sanitize@url \@url }%
\providecommand \@url [1]{\endgroup\@href {#1}{\urlprefix }}%
\providecommand \urlprefix  [0]{URL }%
\providecommand \Eprint [0]{\href }%
\providecommand \doibase [0]{http://dx.doi.org/}%
\providecommand \selectlanguage [0]{\@gobble}%
\providecommand \bibinfo  [0]{\@secondoftwo}%
\providecommand \bibfield  [0]{\@secondoftwo}%
\providecommand \translation [1]{[#1]}%
\providecommand \BibitemOpen [0]{}%
\providecommand \bibitemStop [0]{}%
\providecommand \bibitemNoStop [0]{.\EOS\space}%
\providecommand \EOS [0]{\spacefactor3000\relax}%
\providecommand \BibitemShut  [1]{\csname bibitem#1\endcsname}%
\let\auto@bib@innerbib\@empty
\bibitem [{\citenamefont {Lachance}(2000)}]{LAC00}%
  \BibitemOpen
  \bibfield  {author} {\bibinfo {author} {\bibfnamefont {R.~L.}\ \bibnamefont
  {Lachance}},\ }in\ \href@noop {} {\emph {\bibinfo {booktitle} {Fifth Workshop
  of Infrared Emission Measurements by FTIR}}}\ (\bibinfo {year}
  {2000})\BibitemShut {NoStop}%
\bibitem [{\citenamefont {Chase}(1984)}]{CHA84}%
  \BibitemOpen
  \bibfield  {author} {\bibinfo {author} {\bibfnamefont {D.}~\bibnamefont
  {Chase}},\ }\href@noop {} {\bibfield  {journal} {\bibinfo  {journal} {Applied
  spectroscopy}\ }\textbf {\bibinfo {volume} {38}},\ \bibinfo {pages} {491}
  (\bibinfo {year} {1984})}\BibitemShut {NoStop}%
\bibitem [{\citenamefont {Guelachvili}(1986)}]{GUE86}%
  \BibitemOpen
  \bibfield  {author} {\bibinfo {author} {\bibfnamefont {G.}~\bibnamefont
  {Guelachvili}},\ }\href@noop {} {\bibfield  {journal} {\bibinfo  {journal}
  {Applied optics}\ }\textbf {\bibinfo {volume} {25}},\ \bibinfo {pages} {4644}
  (\bibinfo {year} {1986})}\BibitemShut {NoStop}%
\bibitem [{\citenamefont {Carter~III}\ \emph {et~al.}(1990)\citenamefont
  {Carter~III}, \citenamefont {Lindsay},\ and\ \citenamefont {Beduhn}}]{CAR90}%
  \BibitemOpen
  \bibfield  {author} {\bibinfo {author} {\bibfnamefont {R.}~\bibnamefont
  {Carter~III}}, \bibinfo {author} {\bibfnamefont {N.}~\bibnamefont {Lindsay}},
  \ and\ \bibinfo {author} {\bibfnamefont {D.}~\bibnamefont {Beduhn}},\
  }\href@noop {} {\bibfield  {journal} {\bibinfo  {journal} {Applied
  spectroscopy}\ }\textbf {\bibinfo {volume} {44}},\ \bibinfo {pages} {1147}
  (\bibinfo {year} {1990})}\BibitemShut {NoStop}%
\bibitem [{\citenamefont {Abrams}\ \emph {et~al.}(1994)\citenamefont {Abrams},
  \citenamefont {Toon},\ and\ \citenamefont {Schindler}}]{ABR94}%
  \BibitemOpen
  \bibfield  {author} {\bibinfo {author} {\bibfnamefont {M.~C.}\ \bibnamefont
  {Abrams}}, \bibinfo {author} {\bibfnamefont {G.}~\bibnamefont {Toon}}, \ and\
  \bibinfo {author} {\bibfnamefont {R.}~\bibnamefont {Schindler}},\ }\href@noop
  {} {\bibfield  {journal} {\bibinfo  {journal} {Applied optics}\ }\textbf
  {\bibinfo {volume} {33}},\ \bibinfo {pages} {6307} (\bibinfo {year}
  {1994})}\BibitemShut {NoStop}%
\bibitem [{\citenamefont {Jeseck}\ \emph {et~al.}(1998)\citenamefont {Jeseck},
  \citenamefont {Camy-Peyret}, \citenamefont {Payan},\ and\ \citenamefont
  {Hawat}}]{JES98}%
  \BibitemOpen
  \bibfield  {author} {\bibinfo {author} {\bibfnamefont {P.}~\bibnamefont
  {Jeseck}}, \bibinfo {author} {\bibfnamefont {C.}~\bibnamefont {Camy-Peyret}},
  \bibinfo {author} {\bibfnamefont {S.}~\bibnamefont {Payan}}, \ and\ \bibinfo
  {author} {\bibfnamefont {T.}~\bibnamefont {Hawat}},\ }\href@noop {}
  {\bibfield  {journal} {\bibinfo  {journal} {Applied optics}\ }\textbf
  {\bibinfo {volume} {37}},\ \bibinfo {pages} {6544} (\bibinfo {year}
  {1998})}\BibitemShut {NoStop}%
\bibitem [{\citenamefont {Guay}\ and\ \citenamefont {Genest}(2021)}]{GUA21a}%
  \BibitemOpen
  \bibfield  {author} {\bibinfo {author} {\bibfnamefont {P.}~\bibnamefont
  {Guay}}\ and\ \bibinfo {author} {\bibfnamefont {J.}~\bibnamefont {Genest}},\
  }\href@noop {} {\enquote {\bibinfo {title} {Balanced photodetectors
  nonlinearity for short-pulse regime},}\ } (\bibinfo {year} {2021}),\ \Eprint
  {http://arxiv.org/abs/2105.15192} {arXiv:2105.15192 [physics.ins-det]}
  \BibitemShut {NoStop}%
\bibitem [{\citenamefont {Diddams}\ \emph {et~al.}(2009)\citenamefont
  {Diddams}, \citenamefont {Kirchner}, \citenamefont {Fortier}, \citenamefont
  {Braje}, \citenamefont {Weiner},\ and\ \citenamefont {Hollberg}}]{DID09}%
  \BibitemOpen
  \bibfield  {author} {\bibinfo {author} {\bibfnamefont {S.~A.}\ \bibnamefont
  {Diddams}}, \bibinfo {author} {\bibfnamefont {M.}~\bibnamefont {Kirchner}},
  \bibinfo {author} {\bibfnamefont {T.}~\bibnamefont {Fortier}}, \bibinfo
  {author} {\bibfnamefont {D.}~\bibnamefont {Braje}}, \bibinfo {author}
  {\bibfnamefont {A.}~\bibnamefont {Weiner}}, \ and\ \bibinfo {author}
  {\bibfnamefont {L.}~\bibnamefont {Hollberg}},\ }\href@noop {} {\bibfield
  {journal} {\bibinfo  {journal} {Optics Express}\ }\textbf {\bibinfo {volume}
  {17}},\ \bibinfo {pages} {3331} (\bibinfo {year} {2009})}\BibitemShut
  {NoStop}%
\bibitem [{\citenamefont {Sinclair}\ \emph {et~al.}(2015)\citenamefont
  {Sinclair}, \citenamefont {Desch{\^e}nes}, \citenamefont {Sonderhouse},
  \citenamefont {Swann}, \citenamefont {Khader}, \citenamefont {Baumann},
  \citenamefont {Newbury},\ and\ \citenamefont {Coddington}}]{SIN15}%
  \BibitemOpen
  \bibfield  {author} {\bibinfo {author} {\bibfnamefont {L.~C.}\ \bibnamefont
  {Sinclair}}, \bibinfo {author} {\bibfnamefont {J.-D.}\ \bibnamefont
  {Desch{\^e}nes}}, \bibinfo {author} {\bibfnamefont {L.}~\bibnamefont
  {Sonderhouse}}, \bibinfo {author} {\bibfnamefont {W.~C.}\ \bibnamefont
  {Swann}}, \bibinfo {author} {\bibfnamefont {I.~H.}\ \bibnamefont {Khader}},
  \bibinfo {author} {\bibfnamefont {E.}~\bibnamefont {Baumann}}, \bibinfo
  {author} {\bibfnamefont {N.~R.}\ \bibnamefont {Newbury}}, \ and\ \bibinfo
  {author} {\bibfnamefont {I.}~\bibnamefont {Coddington}},\ }\href@noop {}
  {\bibfield  {journal} {\bibinfo  {journal} {Rev. Sci. Instrum}\ }\textbf
  {\bibinfo {volume} {86}},\ \bibinfo {pages} {081301} (\bibinfo {year}
  {2015})}\BibitemShut {NoStop}%
\bibitem [{\citenamefont {Tourigny-Plante}\ \emph {et~al.}(2020)\citenamefont
  {Tourigny-Plante}, \citenamefont {Guay},\ and\ \citenamefont
  {Genest}}]{TOU20}%
  \BibitemOpen
  \bibfield  {author} {\bibinfo {author} {\bibfnamefont {A.}~\bibnamefont
  {Tourigny-Plante}}, \bibinfo {author} {\bibfnamefont {P.}~\bibnamefont
  {Guay}}, \ and\ \bibinfo {author} {\bibfnamefont {J.}~\bibnamefont
  {Genest}},\ }in\ \href@noop {} {\emph {\bibinfo {booktitle} {Laser
  Applications to Chemical, Security and Environmental Analysis}}}\ (\bibinfo
  {organization} {Optical Society of America},\ \bibinfo {year} {2020})\ pp.\
  \bibinfo {pages} {LTu3C--2}\BibitemShut {NoStop}%
\bibitem [{\citenamefont {H{\'e}bert}\ \emph {et~al.}(2017)\citenamefont
  {H{\'e}bert}, \citenamefont {Genest}, \citenamefont {Desch{\^e}nes},
  \citenamefont {Bergeron}, \citenamefont {Chen}, \citenamefont {Khurmi},\ and\
  \citenamefont {Lancaster}}]{HEB17}%
  \BibitemOpen
  \bibfield  {author} {\bibinfo {author} {\bibfnamefont {N.~B.}\ \bibnamefont
  {H{\'e}bert}}, \bibinfo {author} {\bibfnamefont {J.}~\bibnamefont {Genest}},
  \bibinfo {author} {\bibfnamefont {J.-D.}\ \bibnamefont {Desch{\^e}nes}},
  \bibinfo {author} {\bibfnamefont {H.}~\bibnamefont {Bergeron}}, \bibinfo
  {author} {\bibfnamefont {G.~Y.}\ \bibnamefont {Chen}}, \bibinfo {author}
  {\bibfnamefont {C.}~\bibnamefont {Khurmi}}, \ and\ \bibinfo {author}
  {\bibfnamefont {D.~G.}\ \bibnamefont {Lancaster}},\ }\href@noop {} {\bibfield
   {journal} {\bibinfo  {journal} {Optics express}\ }\textbf {\bibinfo {volume}
  {25}},\ \bibinfo {pages} {8168} (\bibinfo {year} {2017})}\BibitemShut
  {NoStop}%
\bibitem [{\citenamefont {Guay}\ \emph {et~al.}(2018)\citenamefont {Guay},
  \citenamefont {Genest},\ and\ \citenamefont {Fleisher}}]{GUA18}%
  \BibitemOpen
  \bibfield  {author} {\bibinfo {author} {\bibfnamefont {P.}~\bibnamefont
  {Guay}}, \bibinfo {author} {\bibfnamefont {J.}~\bibnamefont {Genest}}, \ and\
  \bibinfo {author} {\bibfnamefont {A.~J.}\ \bibnamefont {Fleisher}},\
  }\href@noop {} {\bibfield  {journal} {\bibinfo  {journal} {Optics letters}\
  }\textbf {\bibinfo {volume} {43}},\ \bibinfo {pages} {1407} (\bibinfo {year}
  {2018})}\BibitemShut {NoStop}%
\bibitem [{\citenamefont {Guay}\ \emph {et~al.}(2020)\citenamefont {Guay},
  \citenamefont {Tourigny-Plante}, \citenamefont {H{\'e}bert}, \citenamefont
  {Michaud-Belleau}, \citenamefont {Larouche}, \citenamefont {Fdil},\ and\
  \citenamefont {Genest}}]{GUA20}%
  \BibitemOpen
  \bibfield  {author} {\bibinfo {author} {\bibfnamefont {P.}~\bibnamefont
  {Guay}}, \bibinfo {author} {\bibfnamefont {A.}~\bibnamefont
  {Tourigny-Plante}}, \bibinfo {author} {\bibfnamefont {N.~B.}\ \bibnamefont
  {H{\'e}bert}}, \bibinfo {author} {\bibfnamefont {V.}~\bibnamefont
  {Michaud-Belleau}}, \bibinfo {author} {\bibfnamefont {S.}~\bibnamefont
  {Larouche}}, \bibinfo {author} {\bibfnamefont {K.}~\bibnamefont {Fdil}}, \
  and\ \bibinfo {author} {\bibfnamefont {J.}~\bibnamefont {Genest}},\
  }\href@noop {} {\bibfield  {journal} {\bibinfo  {journal} {Applied optics}\
  }\textbf {\bibinfo {volume} {59}},\ \bibinfo {pages} {B35} (\bibinfo {year}
  {2020})}\BibitemShut {NoStop}%
\bibitem [{\citenamefont {Larouche}(2020)}]{LAR20}%
  \BibitemOpen
  \bibfield  {author} {\bibinfo {author} {\bibfnamefont {S.}~\bibnamefont
  {Larouche}},\ }\emph {\bibinfo {title} {{Autocorrection en
  interf{\'e}rom{\'e}trie {\`a} double peigne avec deux lasers {\`a} fibres
  ind{\'e}pendants}}},\ \href
  {https://corpus.ulaval.ca/jspui/handle/20.500.11794/6633} {Master's thesis},\
  \bibinfo  {school} {Université Laval} (\bibinfo {year} {2020})\BibitemShut
  {NoStop}%
\bibitem [{\citenamefont {Potvin}\ \emph {et~al.}(2013)\citenamefont {Potvin},
  \citenamefont {Boudreau}, \citenamefont {Desch{\^e}nes},\ and\ \citenamefont
  {Genest}}]{POT13}%
  \BibitemOpen
  \bibfield  {author} {\bibinfo {author} {\bibfnamefont {S.}~\bibnamefont
  {Potvin}}, \bibinfo {author} {\bibfnamefont {S.}~\bibnamefont {Boudreau}},
  \bibinfo {author} {\bibfnamefont {J.-D.}\ \bibnamefont {Desch{\^e}nes}}, \
  and\ \bibinfo {author} {\bibfnamefont {J.}~\bibnamefont {Genest}},\
  }\href@noop {} {\bibfield  {journal} {\bibinfo  {journal} {Applied optics}\
  }\textbf {\bibinfo {volume} {52}},\ \bibinfo {pages} {248} (\bibinfo {year}
  {2013})}\BibitemShut {NoStop}%
\end{thebibliography}%

\end{document}